\documentclass[twoside]{article}
\usepackage[a4paper]{geometry}
\usepackage[utf8]{inputenc} 
\usepackage[OT1]{fontenc}
\usepackage{RR}
\usepackage{hyperref}
\RRNo{8541}
\RRdate{May 2014}
\RRauthor{\L ukasz Domaga{\l}a\thanks{Inria}
\and Fabrice Rastello\thanks{Inria}
\and Sadayappan Ponnuswany\thanks{OSU}
\and Duco van Amstel\thanks{Inria}
}
\authorhead{Domaga{\l}a, Rastello, Ponnuswany and van Amstel}
\RRetitle{A Tiling Perspective for Register Optimization}

\RRtitle{Perspective de tuillage pour l'optimisation de registres.} 

\usepackage{expl3}[2012-07-08]
\ExplSyntaxOn
\cs_new_eq:NN \fpeval \fp_eval:n
\ExplSyntaxOff

\newcommand{\Root}{./}
\newcommand{\FiguresDir}{\Root}

\RRabstract{
Register allocation is a much studied problem. A particularly important context for optimizing
register allocation is within loops, since a significant fraction of the execution time of programs
is often inside loop code. A variety of algorithms have been proposed in the past for register
allocation, but the complexity of the problem has resulted in a decoupling of several important
aspects, including loop unrolling, register promotion, and instruction reordering.

In this paper, we develop an approach to register allocation and promotion in a unified optimization
framework that simultaneously considers the impact of loop unrolling and instruction scheduling.
This is done via a novel instruction tiling approach where instructions within a loop are
represented along one dimension and innermost loop iterations along the other dimension. By
exploiting the regularity along the loop dimension, and imposing essential dependence based
constraints on intra-tile execution order, the problem of optimizing register pressure is cast in a
constraint programming formalism. Experimental results are provided from thousands of innermost
loops extracted from the SPEC benchmarks, demonstrating improvements over the current
state-of-the-art.

}
\RRresume{L'allocation de registres est un problème largement étudié.
Un contexte particulièrement important pour l’optimisation de l’allocation de registres est celui des boucles car elles constituent une fraction importante du temps d'exécution du programme. 
De nombreux algorithmes d'allocation de registres ont été proposés dans le passé mais la complexité du problème à donné lieu à un découplage de plusieurs aspects importants, incluant notamment le déroulage de boucles, la promotion de registres ou le réordonnance d’instructions.

Dans ce rapport nous développons une approche unifiée au problème d'allocation et promotion de registres dans un cadre d’optimisation qui combine l’impact du déroulage de boucles et le réordonnancement d’instructions. 
Ceci est réalisé grâce à une nouvelle approche de pavage-registres dans lequel les instructions du corps de boucle sont représentées le long d’une dimension et les itérations de la boucle interne le long d’une autre dimension. 
En profitant de régularités le long d’une dimension et en imposant à l'ordre intra-tuile les contraintes de dépendances, le problème d’optimisation de la pression registres est exprimée dans un formalisme de programmation par contraintes. 
Les résultats expérimentaux issus de milliers de boucles internes extraites de la suite de benchmarks SPEC, démontrent l’amélioration par rapport à l’état de l’art.} 
\RRkeyword{compilation, compiler optimisation, register allocation,register spilling, register promotion, scheduling, constraint programming, loop transformations, loop unrolling, tiling, register tiling, locality} 
\RRmotcle{compilation, optimisation de compilation, allocation de registres, vidage en mémoire, promotion de registres, ordonnancement, programmation par contraintes, transformation de boucles, déroulage de boucle, découpage de boucles, pavage, localité} 

\RRprojets{GCG}
\URRhoneAlpes

\usepackage{subcaption}
\usepackage{epsfig}
\usepackage{graphics}
\usepackage{amsmath}
\usepackage{amsfonts}
\usepackage{amssymb}

\begin{document}

\makeRR

\section{Introduction}
\label{sec:introduction}
The efficient use of machine registers has been a fundamental compiler
optimization goal for over half a century. Typically, the goal is to minimize
the number of loads (stores) from (to) memory and/or cache.
\emph{Register allocation}~\cite{chaitin:1981:register} is the central technique
and domain of study for this optimizations. Its goal is
to map variables in a program to either machine registers or memory
locations. Register allocation is subdivided into two sub-problems: first, the
\emph{allocation} selects the set of variables that will reside in registers at
each point of the program; then, the \emph{assignment} or \emph{coloring} assigns
each variable to a specific machine register. In general, it is not possible
for all variables of
a program to reside in registers throughout the execution of the program.
The \emph{spilling} problem~\cite{bouchez_spillevery_07} is that of
determining which variables should be stored (or \emph{spilled}) to memory to make
the assignment possible; it aims at minimizing load/store overhead and thus
attempts to maximize the reuse of values held in the
registers.

Register allocation is a very complex problem because of the
interaction of multiple factors. Even within a single basic block, and a fixed
schedule of operations, minimizing the number of loads and stores is very
hard~\cite{liberatore_onlocalra_00}. Further, there are many possible valid
schedules for the order of execution of the instructions and it is generally
recognized that it can have a significant impact on the number of loads/stores:
different orders of instructions
imply different live ranges for values and thus differences in the number of
necessary registers to perform an allocation. For example,
re-materialization~\cite{Briggs:1992:REM:143103.143143,conf/cf/BahiE11}, which 
can be viewed as a form of very limited re-scheduling is the main source of
performance improvement when integrated in the spilling
formulation~\cite{Colombet:2011:SOS:2038698.2038706}. 

A particularly important context for considering the allocation problem is
within loop computations, since they constitute a significant fraction of the
execution time of many codes. In this paper, we take a fresh new look at this
problem of optimizing register usage within innermost loops of programs 
Instead
of the standard approach of analyzing the inter-statement dependencies among the
operations in the loop body and live range of values based on a pre-determined
schedule of
operations, we view the set of operations in a novel two dimensional Cartesian
space that is tiled to optimize register use. This new representation simultaneously
considers both intra-iteration and inter-iteration register reuse, along with
the use of \emph{register promotion}, a technique that enables an inter-iteration
 memory-dependence
flow through a register instead  ~\cite{reg-promotion-PRE}.



Historically, register allocation did not at first consider loops 
when performing the coupled allocation and assignment
problem~\cite{chaitin:1981:register}. Building on the observation that loops
often generate multiple redundant loads of the same memory addresses, new
analytic methods were developed that could detect such unnecessary loads. This
information can then be used for a technique called \emph{scalar
replacement} or \emph{register promotion} that keeps the memory content in a
register until the next read of this value instead of reloading it from
memory~\cite{Callahan:1990:IRA:93542.93553}. Improving on this first solution,
\emph{register pipelining}~\cite{Duesterwald:1992:RPI:647471.727265} provided a
more extended formalization of the reuse problem from the perspective of register
allocation. However, both register promotion and register pipelining do not
consider the possibility for rescheduling the instructions within a loop.

The advent of the Static Single Assignment (SSA) form allowed a new perspective
where allocation could be performed separately from the
assignment~\cite{HGG:2006:RA_SSA}. New aggressive allocation algorithms were
designed that also perform re-materialization to reduce register pressure.
In the context of architectures with ILP (instruction level parallelism), combined 
register allocation and instruction scheduling have also been extensively studied.
Common to both of these approaches targeting the improvement of register reuse
are many orthogonal loop transformations among which is loop unrolling. This
technique, combined with register promotion, allows for a better exposure of register reuse between consecutive loop iterations. 

In this paper, we develop a novel approach to integrated register optimization
considering register pipelining, instruction rescheduling and loop unrolling. 
We note that while these techniques could be extended to also 
address the interplay between register allocation and ILP, it is beyond the
scope and is not addressed by this paper.
Our aim is to show the impact of combining rescheduling and loop unrolling with register allocation to reduce register pressure, hence reducing the amount of shuffle code between memory and registers.
We
integrate the three optimizations into a common framework using our a two-dimensional
tiles loop perspective
and a new model for the computation of spill costs. The
two-dimensional space code is formed by using
one dimension to represent the scheduling of the statement within the inner-most
loop, while the other dimension represents the multiple iterations of this
inner-most loop. This leads to the formulation of an optimization problem in
the constraint programming paradigm. The problem formalizes the optimization of
register reuse by reducing the number of spills and simultaneously considers two
optimizations:
\begin{enumerate}
\item the rescheduling of the instructions in the loop-body
\item the tiling of the two-dimensional loop representation with rectangular
tiles
\end{enumerate}

In the next section, we use a simple example to describe the
approach. We then provide a high-level overview of the steps of the register
optimization algorithm in
Sec.~3. Sec.~4 provides an introduction to the methodology
of constraint programming. Sec.~5 presents the formal description of the
optimization approach. Sec.~6 presents experimental results of
evaluation of our approach, using over 2000 innermost loops extracted from the SPEC
benchmark suite..
Sec.~7 discusses related work and we conclude with a discussion
in Sec.~8.

\section{Motivating Example}
\label{sec:motivation}

\subsection{Presentation of a typical case}

\begin{figure}
\centering
\begin{subfigure}[b]{0.325\textwidth}
\small
\begin{footnotesize}
\begin{tabbing}
{\bf for } $i=0:N-1$\\
\quad $S_0$: $(X[i],\ a,\ d[i][\dots])\gets f_0(X[i-1])$\\
\quad $S_1$: $(c,\ b[i])\gets f_1(a,\ b[i-1])$\\
\quad $S_2$: $(Y[i],\ d[i][\dots],\ e)\gets f_2(Y[i-1])$\\
\quad $S_3$: $f_3(c,\ e)$\\
\end{tabbing}
\end{footnotesize}
\caption{$X[i]$ and $Y[i]$ are structures made of two words each. $a,\ b[i],\ c,\ d[i][*],\ e$ are a
single word each.\label{fig:ToyPseudo}}
\end{subfigure}
\quad
\begin{subfigure}[b]{0.26\textwidth}
\includegraphics[width= 0.8\textwidth]{\FiguresDir/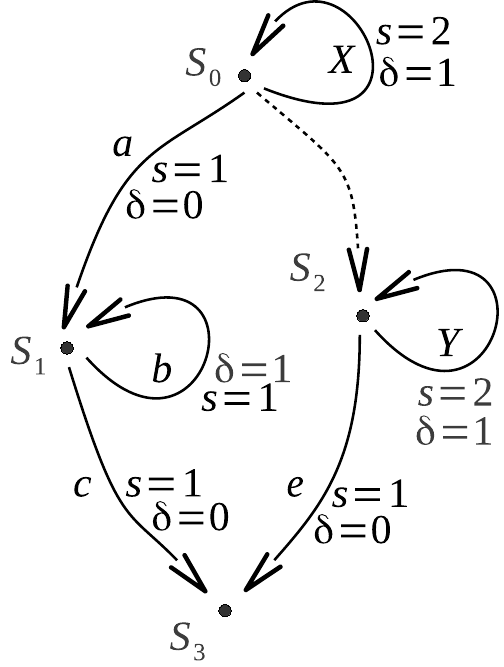}
\caption{$s$ is the size of the flow edge; $\delta$ is the iteration distance.~\\\label{fig:ToyDataFlow}}
\end{subfigure}
\caption{\label{fig:Toy}A toy example and its corresponding dependences. Data-flow must-dependences (candidate for promotion) are represented using solid edges. Other dependences, such as the output may-dependence related to $d$ are represented using dashed edges.}
\end{figure}

Consider the synthetic example given in Figure~\ref{fig:ToyPseudo}. It corresponds to a
counted-loop with induction variable $i$ and a loop-invariant $N$. Its loop body is made of
four statements $S_0$, $S_1$, $S_2$, and $S_3$. In practice, each statement may correspond to the
aggregation of several machine instructions. We refer it as a
\emph{macro-instruction} (see Section~\ref{sec:generalization}). Operands of a macro-instruction can be either scalar or memory variables.
Its semantic is such that input memory variables are first loaded into a register; then computation
is performed atomically using a fixed known number of registers. This number of registers is
referred as the \emph{internal register requirement} of the macro-instruction. Output memory
variables are stored to memory only once computation is fully completed. Clearly the internal
register requirement of a macro-instruction is necessarily greater than or equal to the maximum of
either the sum of the sizes of its inputs or the sum of the sizes of its outputs. As an example, in
statement $S_1(i):~(c,b[i])\gets f_1(a,b[i-1])$, $f_1$ is a macro-instruction that uses the scalar 
variable $a$ and memory variable $b[i-1]$, and that defines the scalar variable $c$ and memory
variable $b[i]$. To illustrate the notion of internal register requirement we suppose its value for
$S_0$, $S_1$, $S_2$, and $S_3$ to be respectively 3, 2, 3, and 2 here. We consider the data-flow
graph of this loop body shown in Figure~\ref{fig:ToyDataFlow}. It has six flow edges that
respectively correspond to memory variables $X[i]$, $Y[i]$, and $b[i]$ and three scalar variables
$a$, $c$, and $e$. All flow edges are labeled with the size of the data they carry (denoted as $s$).
Again for the purpose of illustration we suppose each element of the array, namely $X$, $b$, and $Y$
to be respectively of size 2, 1, and 2; any scalar variable ($a$, $c$, and $e$) is of size 1. All
flow edges are also labeled with their iteration distance (denoted as $\delta$). Here we suppose the flow of any edge to be precisely known
(no may-aliases) and the distance to be constant. This turns out to be the case for our toy example.
Under this condition, all memory flow edges are candidates for flowing through registers instead via
register promotion. Unfortunately, the number of physical registers being limited, the opposite
operation of spilling could also be necessary.

Our goal is to minimize the number of memory-read accesses by also exploiting the effect of
both inter and intra-iteration scheduling on the register pressure. Any edge of our data-flow graph
obviously leads to a dependence edge that constrains the schedule. Those are not the only
dependences. As an example, any anti-dependence, or any flow edge that is not a candidate for
register allocation/promotion such as those with unknown/non-constant distance, or those associated
to a may-alias, will lead to an actual dependence. Also, when the loop body is a macro-instruction
control dependences might be considered. All such additional dependences that constrain the
schedule but not the cost function are represented using dashed edges in the figures. To illustrate
this, we added write accesses in $S_0$ and $S_2$ to the two-dimensional array $d$. Because of the
unknown on the second coordinate, this leads to a may output-dependence from $S_0(i)$ to $S_2(i)$.
Note that the overall resulting dependence graph is acyclic when ignoring self-edges. This is an
important restriction for the formulation of the optimization problem that will be described in
Section~\ref{sec:constraint-prog}. It is also the reason for which the statements of our toy example
turn out to be aggregates of machine instructions : they correspond to strongly connected components
of the original graph (see Section~\ref{sec:generalization} for details).

Suppose we have three registers (in addition to the one used for the loop index). With the given schedule, if not unrolled, there are precisely enough registers to
allocate all variables without the need for spilling. Still, the corresponding code would lead to
loading 5 elements at each iteration: two for $X[i-1]$, one for $b[i-1]$, and two for $Y[i-1]$. 
With register pipelining, sub-scripted variable $b[i]$ can be promoted, saving one memory load per iteration, thus leading to a cost of 4. 
We can do better
by unrolling this loop, rescheduling it, and performing scalar promotion. 
As explained later, with an unrolling factor of 6, and the following schedule $S_0(i),\dots,S_0(i+5), S_2(i), \dots, S_2(i+5), S_1(i), \dots,  S_1(i+2), S_3(i), \dots,  S_3(i+2), S_1(i+3), \dots,  S_1(i+3), S_3(i+3), \dots,  S_3(i+5)$ average cost can be lowered to $18/6=3$ (load of $X[i-1]$ prior to $S_0(i)$ and $Y[i-1]$ prior to $S_2(i)$; load of $a$ and $e$ prior to every $S_1$ and $S_3$; load of $b[i-1]$ prior to $S_1(i)$ and $S_1(i+3)$).
The larger the unrolling
factor, the more opportunities appear for register reuse. However together with the unrolling factor
the optimization problem also grows.

\subsection{A cost-model and its specificities}
\begin{figure}
\centering
\includegraphics[width=0.624\textwidth]{\FiguresDir/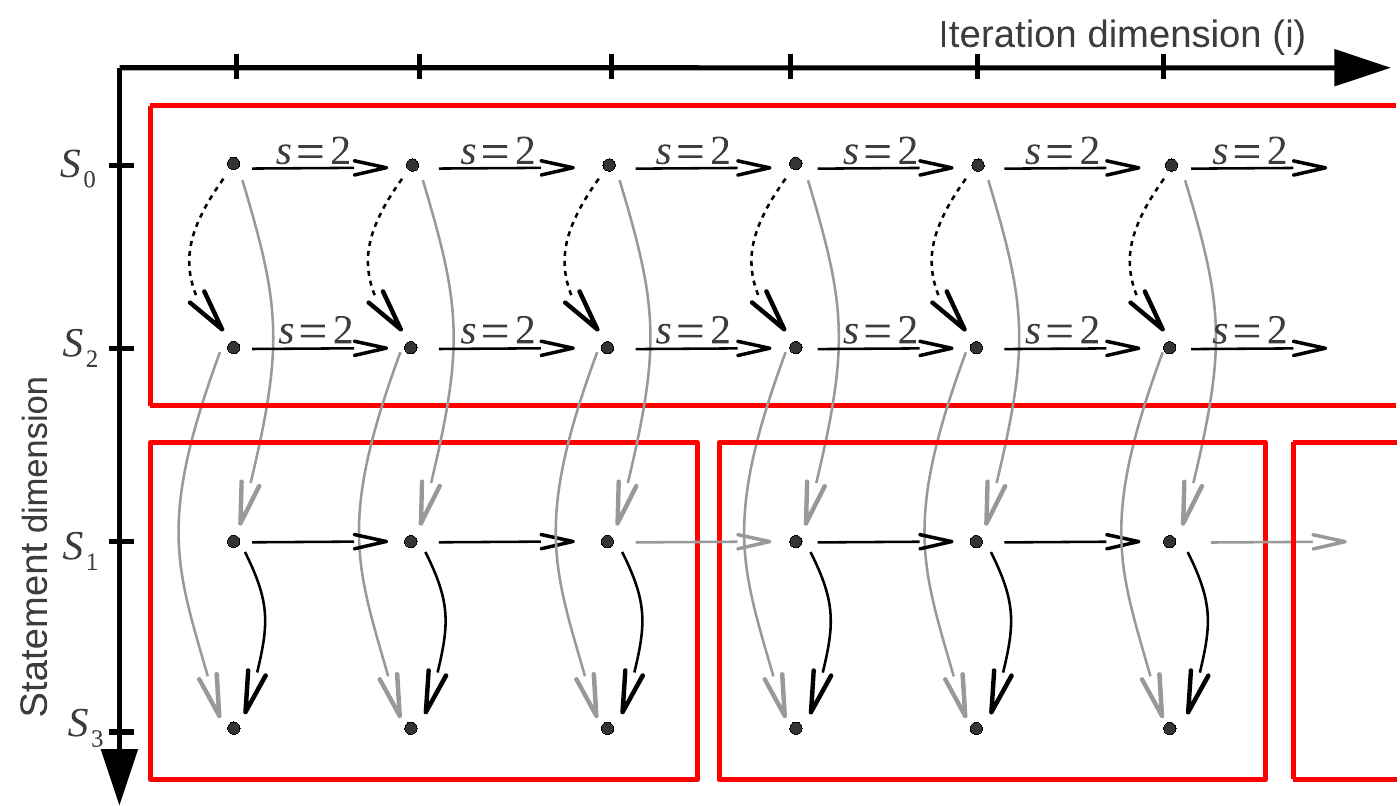}
\caption{\label{fig:ToyTiled} Two-dimensional representation of our problem. Solution expressed as a
tiling of the iteration space. Spilled flow edges are represented in gray. Average spill cost is
$2+1/3$. Value for $s$ is not specified when equal to 1.}
\end{figure}

\begin{figure}
\centering
\includegraphics[width=0.585\textwidth]{\FiguresDir/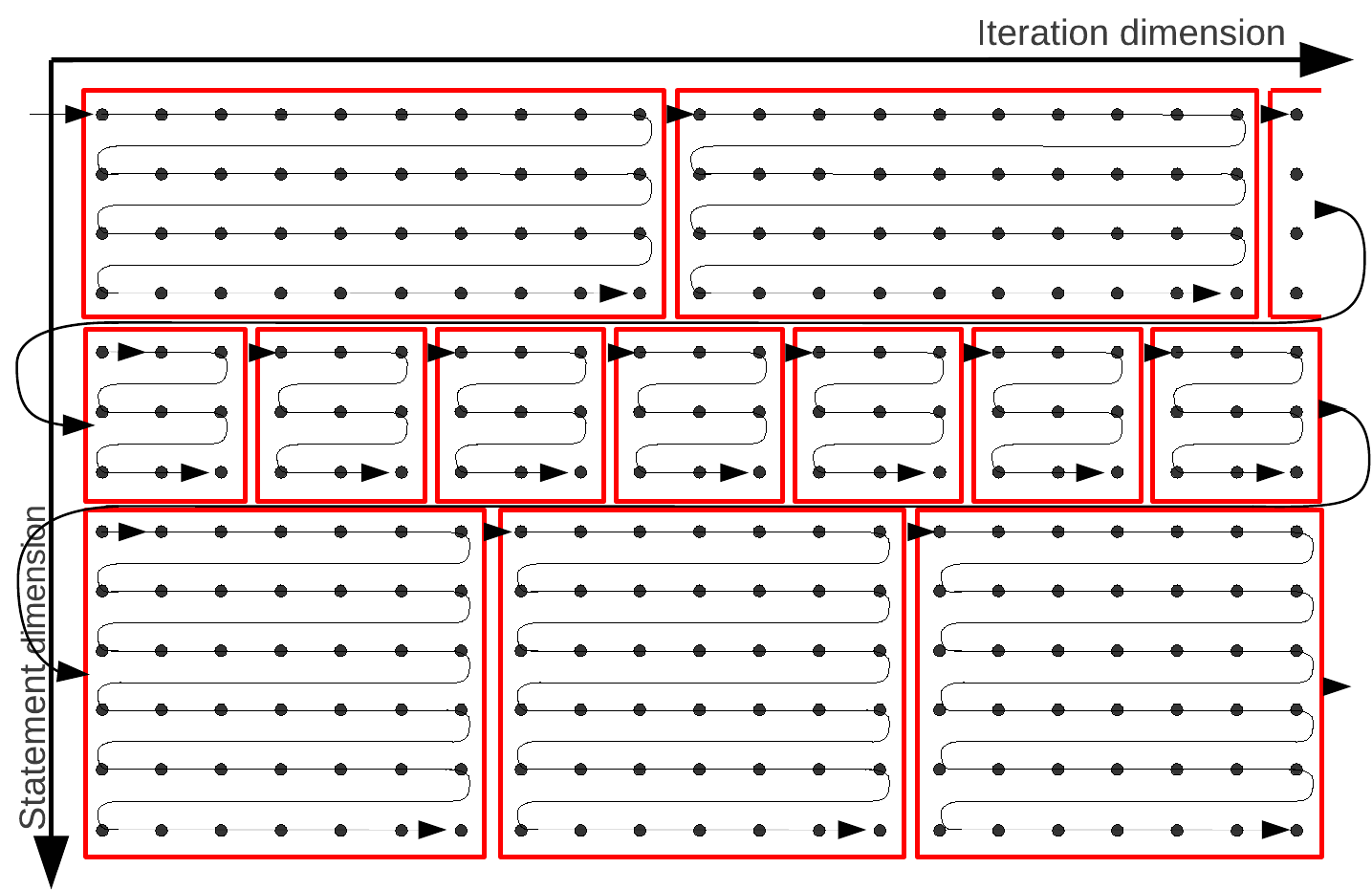}
\caption{\label{fig:ExecOrderNodesAndTiles} A pictorial representation of the overall scheduling of
an iteration space once statements have been linearized.}
\end{figure}

To solve this intrinsically combinatorial problem we represent the loop-body using a two-dimensional
space and use the notion of tiling (see Figure~\ref{fig:ExecOrderNodesAndTiles}) to restrict the
search space and express our solution. As illustrated by Figure~\ref{fig:ToyTiled}, the two
dimensions are respectively the statements (vertically), and the loop iterations (horizontally).
This is nothing else than a view of the data-flow/dependence graph expanded over multiple
iterations. Our solution is equivalent to a register tiling of this iteration space with the
following model:
\begin{itemize}
\item A tile must fulfill the constraint of having a memory requirement that does not
exceed the register capacity. Otherwise put, once the variables that are marked for spilling are out
of consideration the rest of the variables should fit into the available register with any further
need for spilling.
\item Any value that is produced in one tile and consumed during the same iteration (vertical edge) but within another tile is considered to flow through memory (spilled).
\item Any promotable value that is produced in one tile and consumed in a different iteration (horizontal edge) but in the same tile is considered to flow through a register.
\item The register pressure is computed considering that statements within a tile are scheduled row
by row from top to bottom (as illustrated by Figure~\ref{fig:ExecOrderNodesAndTiles}), that loads
are done as late as possible, just before being used, and that stores are done as early as
possible, directly after being produced.
\end{itemize}
The average spill cost, which is to say the number of loads, of our solution for the toy example is
$2+1/3$: $a$ and $e$ are spilled at each iteration and $b[i]$ flows through memory every three
iterations.

\begin{figure}
\centering
\includegraphics[width=0.585\textwidth]{\FiguresDir/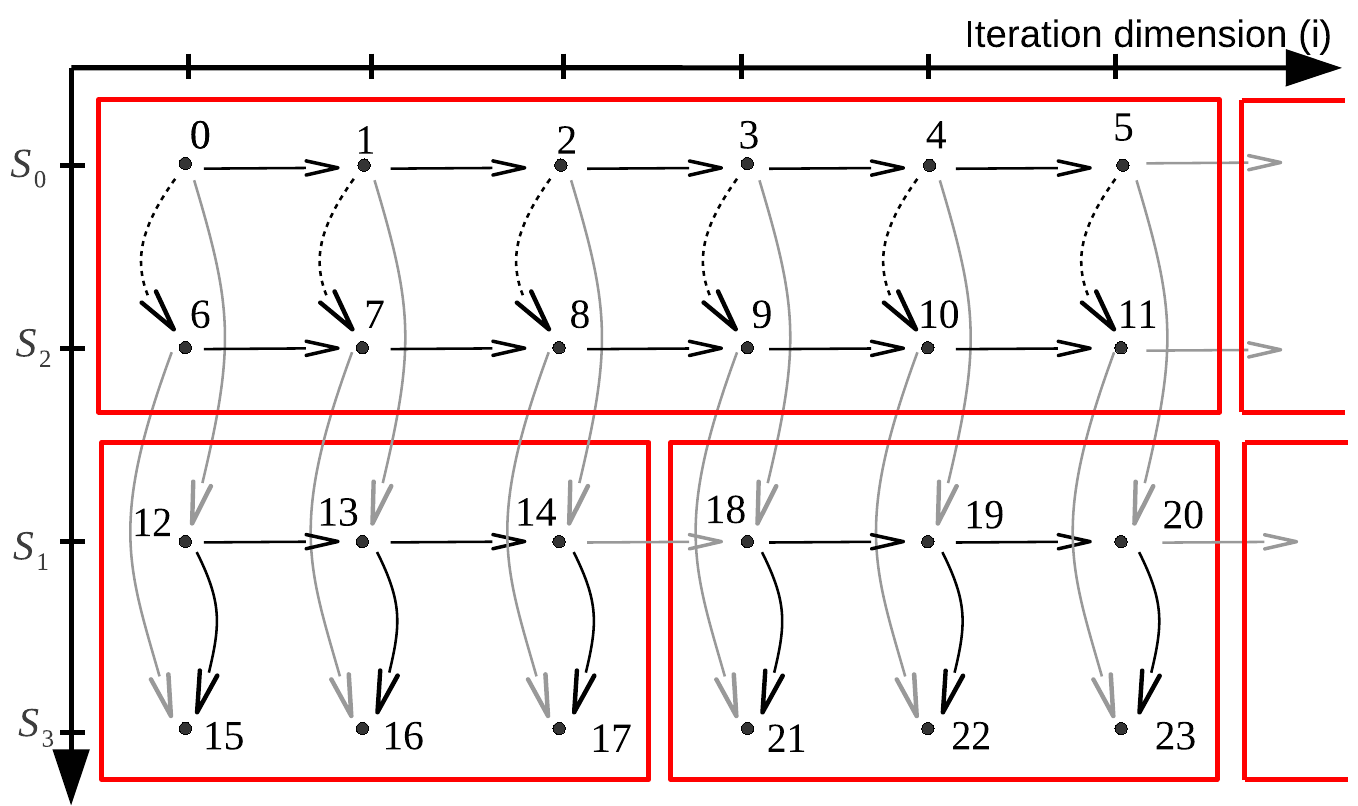}
\caption{\label{fig:ExecOrderUnrolled} Ordering of macro-instructions of the loop body once unrolling with factor 6 have been applied. Average spill cost is $3$.}
\end{figure}

A few points should be noted when considering this approach. First, as opposed to standard register
tiling, the vertical dimension of the iteration space is not uniform and is actually not even a pure
linear dimension. Instead it is a directed acyclic graph instead under one of its linearized forms
and the tiling solution that we seek specifies which valid topological order should be chosen as the
linearization. This is why in our example the obtained ordering of the statements shown by our
solution does not correspond to the initial ordering of the original code. Second, because of the
non-uniformity of this vertical dimension the tiling has no reason to be regular along this
dimension. Recall that the goal of tiling is to expose reuse in both directions and not only
vertically as standard register allocation would do. The shape of each tile is a trade-of between
vertical and horizontal reuse : the wider (respectively higher) the tile, the more we allow for
horizontal (resp. vertical) reuse. This is well illustrated by our example where the upper tile can
be infinitely wide because of the absence of vertical reuse, thus exposing perfect horizontal reuse.
On the opposite, the lower tile width has to be bounded by three because of the presence of
vertical reuse and the row-by-row intra-tile scheduling that is assumed by our model. For this
reason, we look for a tiling scheme that is regular along the horizontal direction but irregular
along the vertical direction. 
Given a set of spilled vertical edges, the tiling can be characterized as follow: (1) a valid linearization of the
statements (topological order); (2) horizontal cuts that define bands; (3) each band
is split in a regular way using vertical cuts; (4) the combination of horizontal
and vertical cuts give the tile shapes.
We should notice that in practice, a maximum allowed unrolling factor is considered as a parameter of our problem statement. This maximum unrolling factor is a trade-of between allowing asymptotic horizontal reuse and limiting code size expansion. 
On our toy example, limiting the unrolling factor to 6 leads to the solution sketched on Figure~\ref{fig:ExecOrderUnrolled} with an average spill cost of 3 instead of the asymptotic cost of $2+1/3$. In this figure, only macro-instructions of one single iteration after unrolling are represented and labels represent the ordering along which they are executed.

\section{A Walkthrough from Source to Binary}
\label{sec:generalization}
\paragraph{Pre-processing} The assumption for applying our optimization approach to a loop is that the loop-body is made out of a single hyperblock. Once this point is satisfied thanks to if-conversion, the rest of our approach can be applied step by step as described in the following section.

\paragraph{Analysis} The first step in our approach is to perform a thorough analysis of the
data-dependences associated with the loop-body and its iterations. The type of analysis may vary
along with the type of the data-structures that are encountered. Some of the possible
data-structures follow:
\begin{itemize}
\item \textbf{Scalars} In the case of scalars, variables that have a single identifiable symbol, it
is necessary to go through the def-use chains in which they are involved. This gives us for each
variable the instructions that assign a value to them and the ones that read this value, resulting
in as many dependence edges as there are def-use pairs.
\item \textbf{Arrays} For arrays we can use Feautrier's Array Dataflow Analysis
method~\cite{Feautrier:2001:ADA:380466.380472}.
\item \textbf{Pointers} The mechanics of pointers and their analysis is a research topic on
itself~\cite{Chase:1990:APS:93542.93585} which looks at the possibility for seperate pointers to
point to the same memory address, a situation that signifies that these pointers are \emph{aliases}.
For our use we are only interested in the most deterministic alias information: the
\emph{must-alias} results. This means that the analyzed pointers are guaranteed to point to the same
memory address, something that is opposed to the \emph{may-alias} result which means that the
aliasing is not guaranteed but still possible.
\end{itemize}

\paragraph{Data-flow graph construction} By gathering all the analysis information that has been
computed during the previous step we construct the data-flow graph of our loop-body. Each observed
dependence results in an edge between the two instructions that are involved. The edge is marked
with the size $s$ of the data targeted by the dependence as well as the dependence distance
$\delta$.

\paragraph{Strongly Connected Components fusion} The obtained data-flow graph may contain Strongly
Connected Components (SCCs). As we want to be able to perform rescheduling on the instructions of
the loop-body it is necessary to consider each SCC as a single macro-instruction. This way the
rescheduling of these macro-instructions will not disturb the execution of their content as it is
considered to be atomical. The new graph is contained by performing a trivial vertice fusion on the
vertices (instructions) of each SCC of the graph. This leaves us with a graph that is now a Directed
Acyclic Graph (DAG) ready to be used for the next step.

\paragraph{Constraint Programming problem writing} Following the details provided in
Sec.~\ref{sec:constraint-prog} it is possible to create a Constraint Programming (CP) instance
corresponding to the optimization problem that we want to solve.

\paragraph{CP instance solving} Using an off-the-shelve CP solver we solve the CP instance corresponding to the current scheduling and tiling problem.

\paragraph{Choice of an unrolling factor} In order to minimize the size of the final code while not
degrading performance further than necessary we search for an unrolling factor that gives a
satisfactory compromise once the final code is generated. The tiles found by the CP solver are only
an indication for scheduling and do not have to be used entirely as will be seen in the next step.

\paragraph{Code generation} The code generation can be divided into several sub-steps:
\begin{enumerate}
\item \textbf{Unrolling} The entire loop-body is unrolled according to the unrolling factor
determined at the previous step. This is done regardless of the result of the tiling.
\item \textbf{Linearization} The unrolled loop-body code is written out tile by tile. Tiles are
processed row by row. The macro-instructions within each tile are also processed row by row. Due to
the fact that the unrolling factor may not be a multiple of the width of the tiles of each row some
tiles may be of reduced width. These are nonetheless processed in the same way.
\item \textbf{Spilling} Loads and stores are inserted for each variable of the code that has not
been retained for register promotion. The insertions are done following the semantics described in
Sec.~\ref{sec:motivation}.
\item \textbf{Register assignment} The remaining variables that are not spilled to memory are
assigned to registers.
\item \textbf{Loop-header initiation} Finally the loop-header is set up and the iteration count is
modified according to the unrolling factor.
\end{enumerate}

\section{Background on Constraint Programming}
\label{sec:cp}
\emph{Constraint programming} \cite{Rossi2006} (CP) is a programming paradigm wherein a program is a
set of \emph{active constraints} over variables with assigned domains and a \emph{search algorithm}.
CP is a \emph{declarative} programming paradigm which means that a program expresses the logic of a
computation without describing its control flow. Within CP:
\begin{itemize}
  \item The \emph{problem model} is a set of variables with assigned domains and active constraints
  over these variables.
  \item A \emph{solution} is an assignment of values to variables from their respective domains; a
  solution is obtained by the search algorithm and \emph{domain filtering}.
  \item A \emph{search algorithm} is an algorithm for assigning values to variables from their
  domains in order to find a solution.
  \item An \emph{active constraint} is a constraint that performs domain filtering (in CP all
  constraints are active and therefore the adjective is omitted in the CP context).
  \item \emph{Domain filtering} is the removal of values from the domains of variables when they are
  not part of any solution.
\end{itemize}

The expressiveness of CP subsumes that of integer linear programming (ILP) which means that all the
problems expressed for ILP can also be solved by CP without any changes to the problem model. In
addition to the linear constraints shared with ILP, CP offers a large portfolio of global
constraints \cite{GlCon2013} such as for example \textit{all different} \cite{Regin1994} and
\textit{global cardinality constraint} \cite{Quimper2004} that encapsulate and ``incrementally
solve'' parts of the problem by performing incremental domain filtering algorithms for global
constraints during the execution of the search algorithm. This removes the need for decomposition
into linear constraints and auxiliary variables which would be otherwise necessary in ILP. 

Constraint programming is a general purpose approach to combinatorial optimization problems that has
been successfully used in different domains such as high level synthesis \cite{Kuchcinski2003},
planning \cite{vanBeek1999}, viechule routing \cite{Shaw1998} or instruction scheduling
\cite{DomagalaThesis2012}. 

The best illustration of the strength of CP is the annual MiniZinc Challenge \cite{Minzinc}, in
which different tools and approaches compete in solving problem from a large benchmark suite. The
challenge with contestants such as ILP solvers (CPLEX, Coin-OR CBC, Gurobi), operations research
tools (Google OR-Tools), CP and CLP solvers (Gecode, JaCoP, Eclipse), CP/SAT solvers (Opturion CPX),
is consistently won by CP solvers. 

One of the biggest advantages of CP, putting aside its extensibility by global constraints, is its
potential for hybridization. This means the cooperation of different tools such as CP solvers, LP
solvers, dynamic programming (DP) or SAT solvers to increase efficiency and reduce search time. Such
hybridization can yield orders of magnitude speed-ups in comparison to the seperate use of these
approaches \cite{Jain2001}\cite{Ottosson2000}\cite{Bollapragada2001}. 

In the case of the tiling problem presented in this paper dynamic programming can solve an instance
where the scheduling of statements is fixed. We thus can build a hybrid CP/DP solver which uses a
dedicated global constraint with a domain filtering algorithm based on DP to reduce search times.
Such a dedicated global constraint has been tested with good results for a simplified tiling problem
as a proof of concept, but integrating it into the current (more detailed) model of the tiling
problem remains an area of future work.

\section{Formalization of Optimization Problem}
\label{sec:constraint-prog}
The entry of our optimization problem is a directed acyclic graph where nodes represent statements
and edges represent dependencies that constrain the scheduling. Dependencies that are associated to
a flow of data that is candidate for register allocation/promotion are labeled with the size of the
data. Any other dependencies are labeled with a fake size of zero. Dependence edges are also
labeled with their iteration distance. Any ``diagonal'' dependence edge, a dependence that is
forward with respect to the loop-body scheduling (i.e $S_a$ to $S_b$) and forward with respect to
the iterations (i.e with a non-zero dependence distance $d>0$), is decomposed into a ``horizontal''
edge (from $S_a$ to itself with distance $d$) followed by a vertical edge (from $S_a$ to $S_b$ with
distance $0$).

The goal of our constraint programming algorithm is to:
\begin{enumerate}
\item find a topological ordering of nodes
\item partition nodes into tiles (nodes within a tile are consecutive according to the ordering)
\item express the register requirement of a given tile as a function of its width, assuming:
  \begin{description}
     \item[\textbf{[schedule]}] nodes are executed row by row from top to bottom 
	 \item[\textbf{[input]}] source of incoming edges are loaded to a register as late as possible
	 right before their first use
	 \item[\textbf{[output]}] source of outgoing edge are stored to memory as soon as they are
	 produced; any register that stores a value is released right after its last use in the tile
     \item[\textbf{[spill-free]}] only first use of input values lead to a load
  \end{description}
\item for each tile, define its width as its largest possible one such that its register requirement
does not exceed the number of available registers
\item express the cost of a tiling as the sum of:
  \begin{itemize}
	\item its \emph{state-cost}: ${\min(\textit{dst, width})\times \textit{state}}/{width}$ for each
	self-edge (horizontal) of size \textit{state} and of distance \textit{dst} for a node that
	belongs to a tile of width \textit{width}
	\item its \emph{stream-cost}: \textit{reg} each time a value of size \textit{reg} is consumed in
	a tile different than the one where it is produced
  \end{itemize}
\item find a tiling with minimum cost
\end{enumerate}

\subsection{Input data}
The input data for the tiling problem is an acyclic graph that comprises:
\begin{itemize}
\item a set of nodes $\{node_i\}_{i \in [0,C]}$, each node with a state corresponding to an inter-iteration self edge of distance one, 
\item a set of directed intra-iteration edges $\{edge_i\}_{i \in [0,E]}$,
\item a set of edge groups $\{group_i\}_{i \in [0,G]}$ such that each edge belongs to exactly one group. An edge group contains edges originating from the same node and corresponding to the same variable. 
\end{itemize}
To create the input data, each original inter-iteration edge had to be decomposed into node state and an intra-iteration edge. The node state of distance greater than one was converted to state with distance one with register requirement multiplied by the original distance.

As part of the input data the aforementioned objects have the following constant properties:
\begin{itemize}
  \item $node_i.state \in \mathbb{N}_0$ - count of registers required to transfer the node state
  between adjacent iterations of a loop,
  \item $node_i.comp \in \mathbb{N}_0$ - count of registers required for execution (internal
  computations) of the node,
  \item $edge_i.src \in \{node_j\}_{j \in [0,C]}$ - source node of the edge,
  \item $edge_i.dst \in \{node_j\}_{j \in [0,C]}$ - destination node of the edge,
  \item $edge_i.reg \in  \mathbb{N}_0$ - count of registers required to transfer data from the
  source to destination node,
  \item $group_i.edge_j \in \{edge_j\}_{j \in [0,E]}$ - $j$-th edge that belongs to the group,
  \item $group_i.reg \in \mathbb{N}_0$ - register consumption of an edge in the group (all are equal within a group).
\end{itemize}

Additionally, as part of the input data, the value of $limit$ is supplied that determines the number
of available registers, and the value of $unoll$ that is the unrolling factor.

\subsection{Problem model}
For the purpose of constructing the problem model we define the following additional objects:

\begin{itemize}
  \item $\{tile_i\}_{i \in [0,C]}$ tiles,
  \item $\{point_i\}_{i \in [0,C]}$ program points.
\end{itemize}

\begin{figure}[h]
\centering
\includegraphics[width=0.585\textwidth]{\FiguresDir/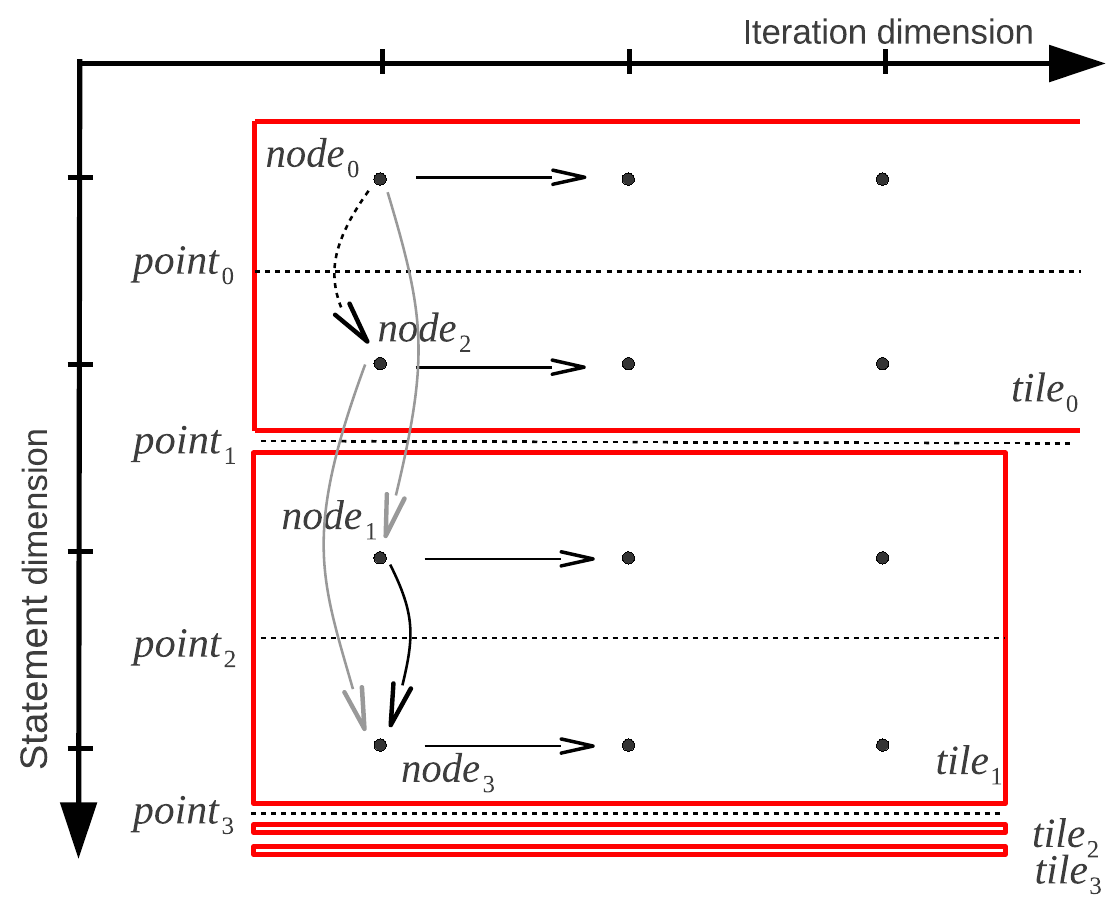}
\caption{\label{fig:CpObjects} Illustration of the CP objects}
\end{figure}

The objects are illustrated in Figure \ref{fig:CpObjects} for the optimal tiling of the motivating
example. Note that $tile_2$ and $tile_3$ exist in the model because the upper bound on the number of
tiles is the count of nodes, but they are not used (do not contain any nodes) in the optimal
solution for this particular problem instance.

\subsubsection{Variables}
We define the following model variables and their domains:

\begin{itemize}
  \item $node_i.order \in [0, C]$ - position of the node in the ordering,
  \item $node_i.tile \in [0, C]$ - index of the tile the node is in,
  \item $node_i.width \in \mathbb{N}$ - width of tile the node is in,
  \item $node_i.spill \in [0, 1]$ - spill the node's last state in the tile
  \item $edge_i.internal \in [0,1]$ - edge does not cross any tile border,
  \item $edge_i.spill \in [0,1]$ - spill the edge,
  \item $edge_i.cross_j \in [0,1]$ - edge crosses point $j \in [0, C]$,
  \item $point_i.tile \in [0, C]$ - index of a tile the point belongs to,
  \item $point_i.press \in \mathbb{N}_0$ - register pressure of the tile at this point,
  \item $point_i.width \in \mathbb{N}$ - width of tile the point belongs to,
  \item $point_i.comp \in \mathbb{N}_0$ - register pressure of a node before this point,
  \item $tile_i.point \in [0, C]$ - point where the tile border is placed,
  \item $tile_i.width \in \mathbb{N}$ - width of the tile,
  \item $group_i.cross_j \in [0,1]$ - reflects whether any of the edges belonging to the group is
  internal and traverses a point $j \in [0, C]$.
\end{itemize}

All the variables' domains are discrete. A variable with domain $[0,1]$ is considered a boolean for
which 0 corresponds to $false$ and 1 to $true$.

\subsubsection{Constraints}
We impose the following constraints:

\begin{align}
&alldifferent(node_0.order,...,node_C.order) \label{equ:node-aldiff}
\end{align}

that ensures each node has a unique index in the ordering,

\begin{align}
&\forall i: \forall j: (tile_{j-1}.point < node_i.order < tile_j.point) \Leftrightarrow
\nonumber\\&\qquad ( node_i.tile = j \wedge node_i.width = tile_j.width )\label{equ:node-tile}
\end{align}

that ensure the assignment of a node to a tile,

\begin{align}
&\forall i: edge_i.src.order < edge_i.dst.order \label{equ:edge-node-order}
\end{align}

that ensure the ordering of nodes imposed by the edges, 

\begin{align}
&\forall i: edge_i.internal \Leftrightarrow (edge_i.src.tile =
edge_i.dst.tile)\label{equ:edge-internal}
\end{align}

that defines an edge as internal if its source and destination node are in the same tile,

\begin{align}
&\forall i: edge_i.spill \geq \neg edge_i.internal  \label{equ:edge-spill}
\end{align}

that ensures all edges that cross tile borders are spilled (leaving the choice to spill or not the internal edges),

\begin{align}
&\forall i: \forall j: edge_i.cross_j \Leftrightarrow \nonumber\\&\qquad(edge_i.src.order \leq j <
edge_i.dst.order) \label{equ:edge-point-cross}
\end{align}

meaning that an edge crosses point $j$ if its source node is before the point and destination node
is after,

\begin{align}
&tile_{-1}.point = -1, tile_C.point =  C \label{equ:tile-border-cond}
\end{align}

defining the tile borders of the first and last tile, tile $tile_{-1}$ does not exist but its
$point$ property is defined as a constant for the purpose of the following constraints,

\begin{align} 
&\forall i: tile_{i-1}.point \leq tile_i.point \label{equ:tile-orderd}
\end{align}

meaning that tiles are ordered according to their indexes, and the end tile border of one tile is
the start tile border of the following tile. A tile $i$ is not used (does not contain any nodes) if
$tile[i-1].point=tile_i.point$,

\begin{align}
&\forall i: \forall j: group_i.cross_j  \Leftrightarrow (\exists group_i.edge_{k-1} \Rightarrow
\nonumber\\&\qquad  (group_i.edge_{k-1}.internal \wedge group_i.edge_{k-1}.cross_j = 1 )
\end{align}

meaning that an edge group crosses a point if any of the edges belonging to the group is internal
and crosses the point,

\begin{align}
&\forall i: (point_i.comp = node_j.comp)  \Leftrightarrow (node_j.order = i)
\end{align}

reflecting internal register usage of a node ordered immediately before the point,

\begin{align}
reserve = \sum_i node_i.state* \neg node_i.spill
\end{align}

defining $reserve$ as the sum of all the states that cross tile border and have not been spilled,

\begin{align}
&\forall i: point_i.press = point_i.comp + reserve +\nonumber\\
&(\sum_j group_j.cross_i \cdot group_j.reg \cdot point_i.width)
\end{align}

meaning that the register pressure in a tile at a point is equal internal register
consumption of a node immediately preceding the point, plus the sum of the border crossing nodes' states that were not spilled, plus the register usage of edge groups
crossing the point, scaled by the width of the tile the point belongs to,

\begin{align}
&\forall i: point_i.press \leq limit
\end{align}

register pressure at each point has to be smaller or equal to the number of available registers.

The $global\_cardinality\_constraint$ \cite{Regin1996} is used to break symmetry in the problem model (forbid equivalent solutions) by ensuring that not used tiles (tiles that don't contain any nodes) only occur at the last position in the ordering (rather than at any position):
\begin{align}
&global\_cardinality\_constraint(\nonumber\\
&\quad(tile_0.point,...,tile_{C-1}.point,tile_C.point),\nonumber\\
&\quad(\{0,1\},...,\{0,1\},\{0,C-1\})\nonumber\\
&)
\end{align}
Breaking symmetry is not necessary for the correctness of the model, but is desired because it speeds-up the search for an optimal solution.

\subsubsection{Cost function}

The cost is represented by the variable $uspill$ that reflects the amount of spill (of the state and
dependence edges) of the unrolled loop.

\begin{align}
uspill = &\sum_i edge_i.reg  \cdot edge_i.spill \cdot  unroll  + \nonumber\\ 
&\sum_j \Big\lceil \frac{unroll}{node_j.width} \Big\rceil  \cdot node_j.state \cdot node_j.spill
\end{align}

The first component of $uspill$ is the cost of all spilled edges. The
second component is the sum of not spilled states of nodes that cross tile borders. The ceiling function computes how many repetitions of the tile (that the node belongs to) fit (entirely or partially) in the unrolled loop.

\begin{align}
spill =\frac{ uspill}{unroll}
\end{align}

An auxiliary variable $spill$ represents the amount of spill normalized for a single iteration of the loop. 

\begin{align}
\min(spill)
\end{align} 

The optimization objective is the minimization of the spill.

\subsubsection{Control variables}
The control variables of a CP problem are a subset of the variables such that when they are assigned
values, all other model variables, including the cost variable, automatically receive values (through
domain filtering of the active constraints). In our tiling problem we are concerned with
ordering the nodes, assigning them to tiles of particular height and width, and make the decisions about spilling internal edges and states, therefore the control variables
are:

\begin{align}
\{&node_0.order,...,node_C.order,\nonumber\\
&tile_0.point,...,tile_C.point,\nonumber\\
&tile_0.width,...,tile_C.width\nonumber\\
&edge_0.spill,...,edge_E.spill,\nonumber\\
&node_0.spill,...,node_C.spill\nonumber\}
\end{align}

\subsection{Search algorithm}
The search algorithm used is the depth-first search with 2-way branching. The variable choice
strategy is "most-constrained" in its dynamic version, meaning that variables involved in the
largest number of constraints are chosen first for assignment of a value and that the number of
constraints is monitored dynamically during the search. The values for the variables are chosen
according to different strategies: for $node_i.order$ values are tested in random order, for
$tile_i.point$ maximal value in the domain is tested first, for $tile_i.width$ the maximal value in
the domain is tested first, for $edge_i.spill$ and $node_i.spill$ the minimal value in
the domain is tested first.

\subsection{CP system}
The constraint programming system chosen to implement the model, search algorithm and custom constraints was JaCoP, due to its large portfolio of global constraints and good results in the MiniZinc Challenges.

\section{Experimental Evaluation}
\label{sec:experiments}
We evaluated our approach on a set of kernels namely \texttt{dct, latanal,  shellsort,} and \texttt{strtrim} and C benchmarks (\texttt{gcc, gzip, crafty, twolf, bzip2, mcf, vpr}) extracted from the specCint2000 suite. Our algorithm has been
implemented in the Open64 compiler with back-end for x86. Scalar replacement~\cite{Callahan:1990:IRA:93542.93553} is aggressively applied by default at early stages of the compiler. The analysis part described in Section~\ref{sec:generalization} whose goal is (in addition to compute all dependences) the detection of must data-flow dependences with the precise corresponding distance $\delta$, is implemented as a small extension of the alias and dependence analysis already available in the compiler: precise dependence and alias information that are used for software pipelining is available at the back-end level for all reducible innermost loops. Hyper-block formation, already available in our branch of the Open64 compiler, has been enabled. 
As a baseline for comparison, we implemented register pipelining technique developed in~\cite{Duesterwald:1992:RPI:647471.727265}.

\paragraph{Interesting Loops}
To evaluate the potential of improvement, we previously performed some statistics using \texttt{gcc} on: (1) the number of strongly connected components (SCCs) of the DAG obtained after SCC fusion described in Section~\ref{sec:generalization}, and (2) on the register pressure.

\begin{figure}[h]
\centering
\includegraphics[scale=0.65]{\FiguresDir/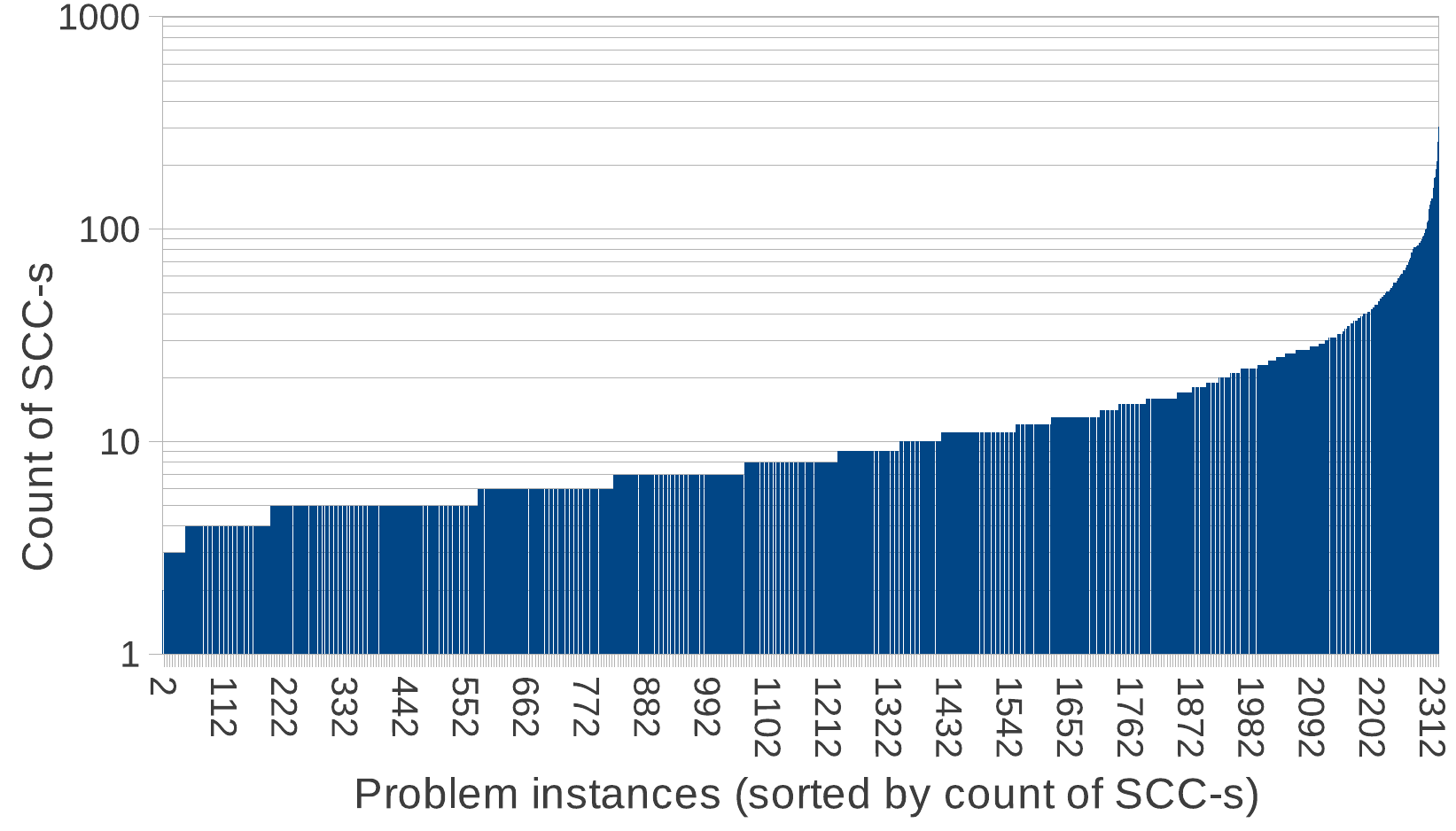}
\caption{\label{fig:SccCnt} Number of strongly connected components (SCCs) for each of the 2327 problem
instances extracted from \texttt{gcc}. The more SCCs a loop has the more potential there is for optimization by rescheduling.}
\end{figure}

\begin{figure}[h]
\centering
\includegraphics[scale=0.65]{\FiguresDir/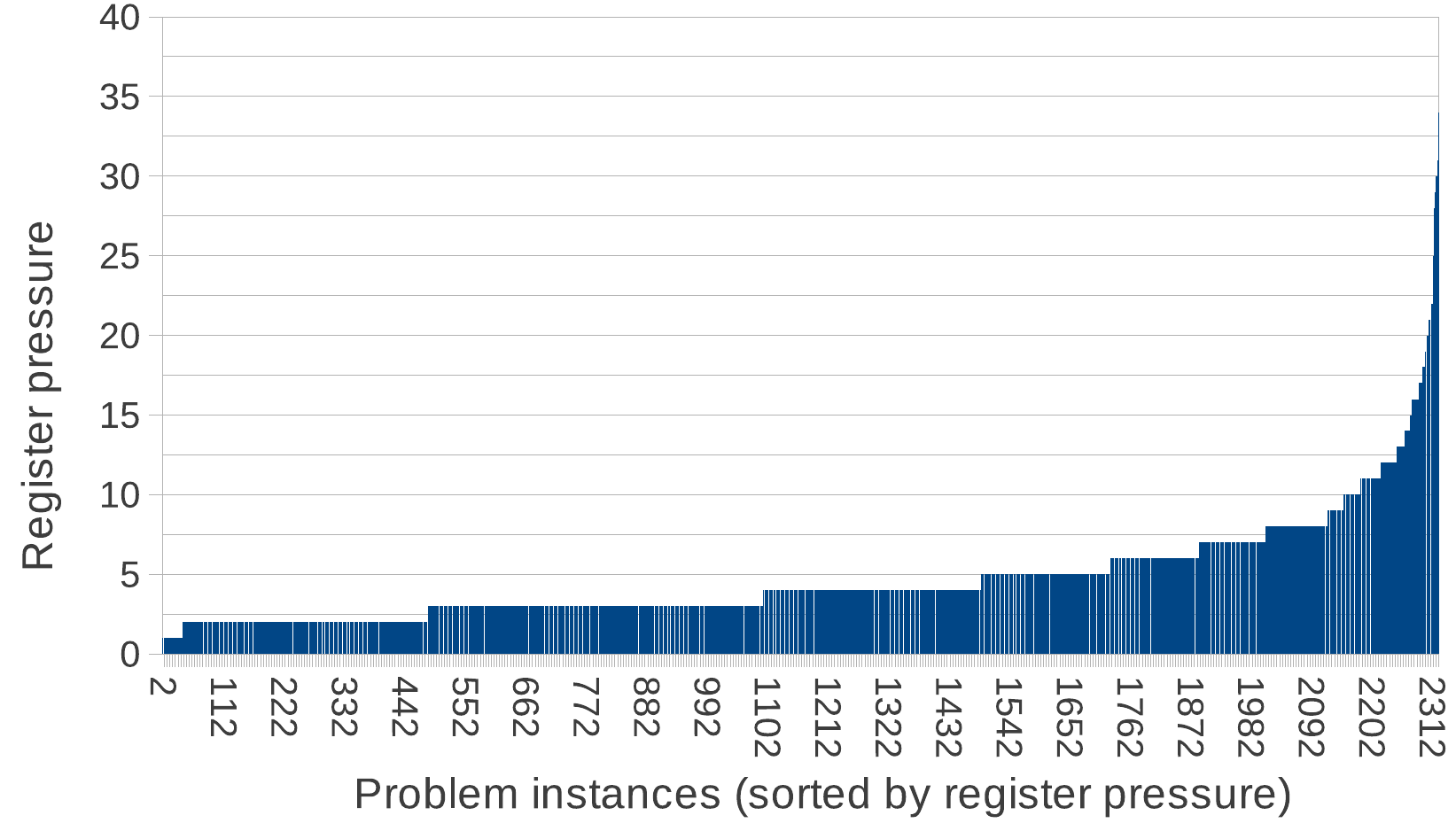}
\caption{\label{fig:RegPress} Maximal register pressure for each of the 2327 problem instances.}
\end{figure}

For each of the loops the program dependence graph has been computed, leading to the aggregation of
instructions into nodes based on strongly connected components (SCCs). The potential for improvement
grows with the number of SCCs as this indicates the freedom for scheduling. For this reason it is
interesting to have statistics on the typical number of SCCs that were found in the loops.
Figure~\ref{fig:SccCnt} reports this number for each instance. Once instructions have been
aggregated into nodes, the resulting dependence graph is acyclic (DAG). For this DAG, the data-flow
graph has been computed: (1) any def-use chain that corresponds to a scalar variable leads to a
data-flow edge; (2) any flow must-dependence for which the distance is a constant also leads to a
data-flow edge. For the statement schedule resulting from the original code, in a very similar way than performed in~\cite{Duesterwald:1992:RPI:647471.727265}, live-ranges and
register pressure were computed based on this data-flow graph. 
Loops for which register pressure is lower than the number of
available physical registers, the reuse cannot be improved by any loop transformations at the
innermost loop level, instead loop interchange or register tiling would be necessary in such cases.
Figure~\ref{fig:RegPress} reports computed register pressure for each problem instance. 

\begin{figure}
\centering
\includegraphics[scale=0.65]{\FiguresDir/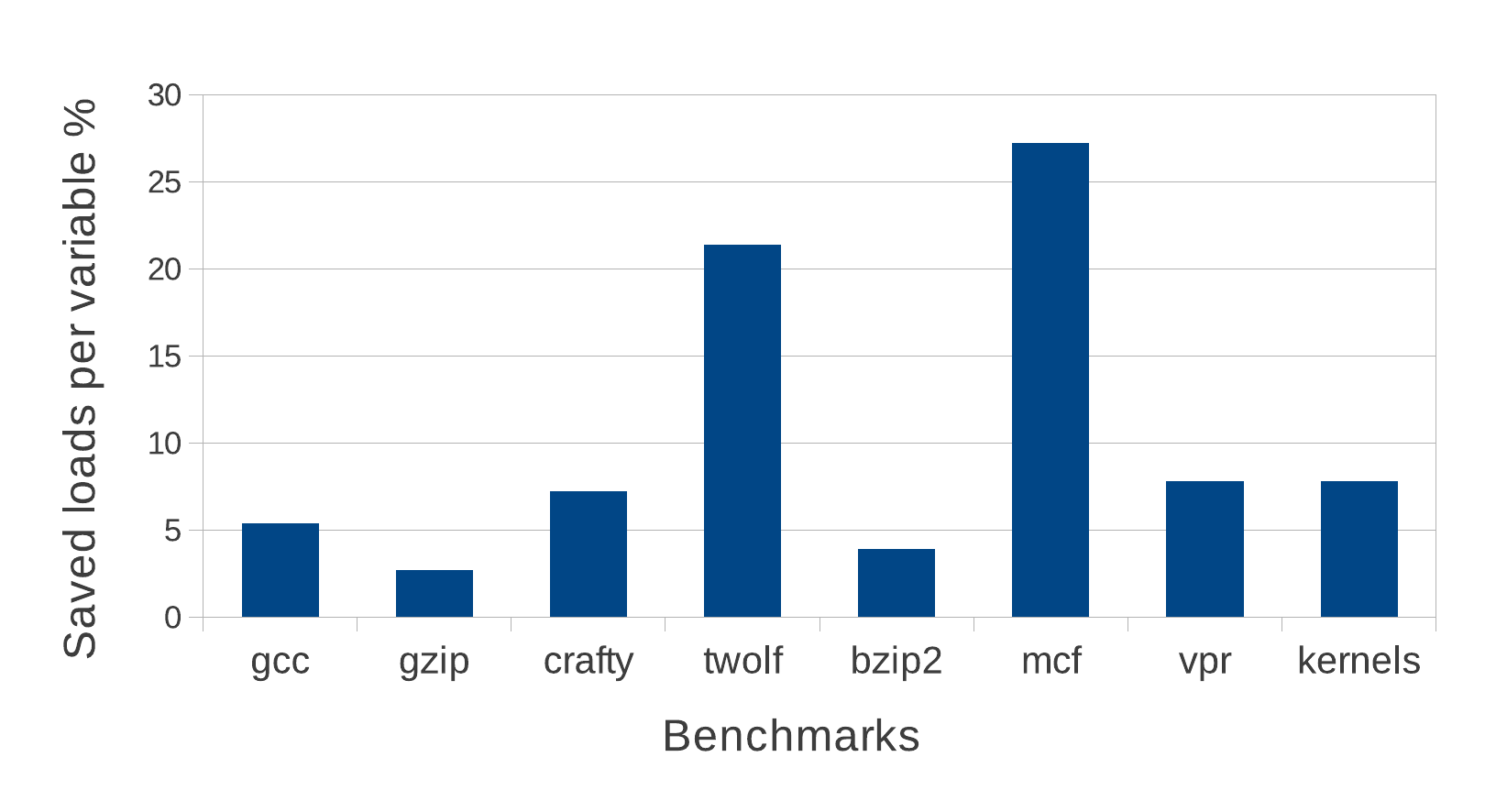}
\caption{\label{fig:LoadSave} Percentage of saved load by combining re-scheduling and unrolling to scalar replacement compared to the baseline (register pipelining). Results are normalized by the number of variables.}
\end{figure}

\paragraph{Spill Cost Improvement}
To evaluate our approach, we counted for each loop of interest: \begin{enumerate}
\item $\#vars$: the number of variables live in the hyper-block;
\item $\#\textit{load\_base}$: the number of load instructions generated by the baseline approach i.e. register pipelining without combining with unrolling and re-scheduling; 
\item $\#\textit{load\_cp}$: the number of load instructions generated by the solution obtained using our constrained programming
\end{enumerate}
Because not all possible schedules are considered by our modeling, there are cases where the constrained programming will not find better solution than the baseline. We compute $\#\textit{load}=\min(\#\textit{load\_base},\#\textit{load\_cp})$. The reported numbers in Figure~\ref{fig:LoadSave} are $100\times\frac{\#\textit{load}-\#\textit{load\_base}}{\#\textit{vars}}$.

\section{Related Work}
\label{sec:related}
\paragraph{Register allocation and scheduling}
The optimal register allocation problem is NP-complete~\cite{{chaitin:1981:register}}. Given an
instruction schedule, good
heuristics~\cite{DBLP:journals/toplas/GeorgeA96,Chow:2004:RAP:989393.989406} or ``optimal''
formulations~\cite{Appel00optimalspilling,Colombet:2011:SOS:2038698.2038706} have been developed for
register allocation to minimize spills. However, there is a strong interaction between the
scheduling of instructions and optimizing register allocation. An integrated optimization of
instruction scheduling and register allocation was shown to be NP-hard by Motwani et
al.~\cite{motwani1995combining}, who then developed a weighted heuristic that allowed control on the
relative priority given to instruction level parallelism versus register pressure.

\paragraph{Re-materialization}
Re-materialization~\cite{{Briggs:1992:REM:143103.143143},{conf/cf/BahiE11}} involves the
regeneration of values from available variables in registers instead of spilling a value to memory
and reloading it. It can be viewed as a form of a very limited re-scheduling and is the main source
of performance when integrated in the spilling formulation~\cite{Colombet:2011:SOS:2038698.2038706}.

\paragraph{Register Tiling}
Register tiling~\cite{DBLP:conf/lcpc/RenganarayanaRR05} considers perfectly nested multi-dimensional
loops with uniform dependencies and represents the innermost loop body as an atomic unique
instruction. These restrictions allow for the optimization of register reuse across multiple loop
dimensions. In this paper we restrict ourselves to the inner-most loop level but expose register
reuse inside the loop body itself. Although unroll and Jam~\cite{Ma:2007:RPG:1369337} can be seen as
a special case of register tiling with rectangular tiles these two techniques are viewed quite
differently in different communities. Unroll and jam and loop unrolling are typically used as means
of expanding the number of statements in the loop body so as to increase instruction level
parallelism.

\section{Discussion and Conclusion}
\label{sec:conclusion}
In this paper, we have re-examined the much studied problem of register optimization using a novel
new approach. We view the set of instructions executed in a loop as a two-dimensional iteration
space, and the register optimization problem as that of finding optimal tile sizes to minimize
spilling. A constraint programming formalism was used to solve the optimization problem. 
The comparison toward the state of the art register pipelining solution, shows in terms of loads instructions substantial improvements.

\bibliographystyle{plain}
\bibliography{\Root/tiling}

\end{document}